\documentclass{article}
\usepackage{spconf,amsmath,graphicx, hyperref,xcolor,acronym}
\usepackage{caption} 
\usepackage{colortbl}
\definecolor{lightgray}{gray}{0.8}

\acrodef{STFT}{short-time Fourier transform}
\acrodef{LSTM}{long short-term memory}
\acrodef{STFTM}{short-time Fourier transform magnitude}
\acrodef{GRU}{gated recurrent unit}
\acrodef{DNNs}{deep neural networks}
\acrodef{MSE}{mean-squared error}
\acrodef{RNN}{recurrent neural network}
\acrodef{LPS}{log-power spectra}
\acrodef{SNR}{signal-to-noise ratio}
\acrodef{PESQ}{perceptual evaluation of speech quality}
\acrodef{STOI}{short-time objective intelligibility}
\acrodef{CD}{cepstral distance}
\acrodef{SI-SDR}{scale-invariant signal-to-distortion ratio}
\acrodef{MOS}{mean opinion score}
\acrodef{FD}{frequency-dependent}
\acrodef{FI}{frequency-independent}

\usepackage{flushend}
\title{Weighted speech distortion losses for \\neural-network-based real-time speech enhancement}
\name{\parbox{\linewidth}{\centering Yangyang Xia$^{1}$\thanks{Yangyang performed this work as an intern while at Microsoft Research.}, Sebastian Braun$^{2}$, Chandan K. A. Reddy$^{2}$,\linebreak Harishchandra Dubey$^{2}$, Ross Cutler$^{2}$, Ivan Tashev$^{2}$}}
\address{$^1$ Dept. of Electrical and Computer Engineering, Carnegie Mellon University, Pittsburgh, PA, USA\\
         $^2$ Microsoft Corporation, Redmond, WA, USA\\
          \href{mailto:raymondxia@cmu.edu}{\texttt{raymondxia@cmu.edu}}, \texttt{\{firstname.lastname\}@microsoft.com}
}

\begin{document}
\ninept
\maketitle
\begin{abstract}
This paper investigates several aspects of training a RNN (recurrent neural network) that impact the objective and subjective quality of enhanced speech for real-time single-channel speech enhancement. Specifically, we focus on a RNN that enhances short-time speech spectra on a single-frame-in, single-frame-out basis, a framework adopted by most classical signal processing methods. We propose two novel mean-squared-error-based learning objectives that enable separate control over the importance of speech distortion versus noise reduction. The proposed loss functions are evaluated by widely accepted objective quality and intelligibility measures and compared to other competitive online methods. In addition, we study the impact of feature normalization and varying batch sequence lengths on the objective quality of enhanced speech. Finally, we show subjective ratings for the proposed approach and a state-of-the-art real-time RNN-based method.
\end{abstract}
\begin{keywords}
Real-time speech enhancement, recurrent neural networks, loss function, speech distortion, mean opinion score
\end{keywords}
%
\section{Introduction}
\label{sec:intro}
%
Speech enhancement (SE) algorithms aim at improving speech quality and intelligibility of speech signals degraded by additive noise \cite{loizou2013speech}, in order to improve human or machine interpretation of speech. Typical SE applications are hearing aids, automatic speech recognition, and audio/video communications in noisy environments. Most SE methods apply a spectral suppression gain or filter to the noisy speech signal in a time-frequency domain \cite{Benesty2005}. In recent supervised learning methods using \ac{DNNs}, a DNN is typically set up to estimate this time-varying gain function \cite{wang2014training} from one or more sets of features derived from noisy speech.

Online processing capability is an attractive feature of a SE algorithm and required for real-time communication applications. Although most classical SE methods have to accommodate their approaches \cite{cohen2002noise,cohen2001speech,ephraim1984speech,boll1979suppression} for fulfilling causality, many DNN-based methods in the literature \cite{wang2014training,ephrat2018looking,Pascual2017} do not enforce this constraint. Several DNN-based approaches report high-quality enhancement using generous look-ahead \cite{ephrat2018looking,Pascual2017}, but their performance for decreasing look-ahead is not well investigated. Nevertheless, DNN-based systems are preferred over classical methods for their ability to accurately suppress transient noise. In this work, we study real-time speech enhancement with \ac{RNN}. Recent works involving RNNs demonstrated promising results \cite{valin2018hybrid}, even at very low \ac{SNR} scenarios \cite{tan2018convolutional,xia2018priori}.

A key challenge in designing a SE algorithm for audio/video communication is to preserve perceived (subjective) speech quality to the best extent possible while suppressing the noise. In classical literature, optimizing such a compound global objective can be done by solving a constrained objective function \cite{braun2015}. Alternatively, one can optimize a simpler objective such as the (log) \ac{MSE} \cite{ephraim1984speech,ephraim1985speech} and employ post-processing modules such as residual noise removal \cite{boll1979suppression} and gain limiting \cite{esch2009}. By contrast, one major benefit of the deep learning framework is the relative ease to incorporate complex learning objectives one believes would drive the enhanced speech towards better quality and intelligibility. Methods along this line of thought include learning multiple objectives from heterogeneous features \cite{sun2017multiple,xu2015multi,Germain2019}, jointly optimizing the final goal and its sub-targets (\emph{e.g.} speech-presence probability) \cite{valin2018hybrid,xia2018priori}, and directly optimizing towards an objective measure of speech quality or intelligibility \cite{martin2018deep,zhao2018perceptually}. The latter seems a promising way to improve objective quality, although both models have to incorporate the standard \ac{MSE} due to the band limitation of each objective measure. \cite{Kumar+2016} reported that a simple perceptually weighted wide-band \ac{MSE} alone does not improve objective speech quality or intelligibility, suggesting that the \ac{MSE} is still a reliable learning objective for wide-band speech enhancement.

In this paper, we propose a DNN-based online speech enhancement system for real-time applications. First, we will discuss features and normalization techniques that would facilitate pattern learning with a \ac{RNN}. We then describe a compact \ac{RNN} that produces a gain function from a single noisy frame. Next, we introduce two simple \ac{MSE}-based loss functions with separate control of speech distortion and noise reduction. During the evaluation, we thoroughly examine the effect of error weighting on the subjective and objective speech quality and intelligibility measures. Furthermore, we discuss how the objective metrics are affected by different feature normalization techniques and training strategies.
%
\section{Problem formulation}
\vspace{-1mm}
\label{sec:problem}
We assume the microphone signals to be described in the \ac{STFT} domain by
\begin{equation}
  X[t,k] = S[t,k] + N[t,k],
\end{equation}
where $X[t,k]$, $S[t,k]$, and $N[t,k]$ denote the \ac{STFT} at time frame $t$ and frequency bin $k$ of the observed noisy speech, clean speech, and noise, respectively. Our system seeks for a time-varying gain function in the \ac{STFTM} domain, $G[t,k]$, that recovers $|S[t,k]|$ to the best extent possible.
\begin{equation}
|\hat{S}[t,k]| = G[t,k]|X[t,k]|
\end{equation}
In real-time processing, $G[t,k]$ shall depend only on the past and present information of input, and is given by
\begin{equation}
G[t,k] = n(g(f(|X[l,k]|));\Theta), l \leq t,
\end{equation}
where $f$ is a transform function applied on the \ac{STFTM} of the noisy signal, $g$ is a normalization function, and $n$ is a DNN whose adaptable parameters are together denoted by $\Theta$. Finally, noisy phase of $X[t,k]$ is applied to $|\hat{S}[t,k]|$ to obtain the enhanced signal.

In the following sections, we will review the state-of-the-art methods, before discussing our choices for $f$ and $g$, the architecture of $n$, two learning objectives for $\Theta$, and further considerations in training that we believe will impact the quality of enhanced speech.

\section{State-of-the-art online noise reduction}
\label{sec:sota}

Classical online SE methods typically seek for the optimal gain function by optimizing some objective functions in a statistical sense. One of the most effective methods in this category assumes the clean and noise \ac{STFT} are uncorrelated, complex Gaussian distributions, and solves for $G[t,k]$ by minimizing the \ac{MSE} between clean and enhanced \ac{STFTM} \cite{ephraim1984speech} or log-\ac{STFTM} \cite{ephraim1985speech}. Although more advanced noise and speech-presence probability models could be incorporated to improve speech quality and prevent musical noise \cite{cohen2002noise, cohen2001speech}, retaining speech quality while removing highly non-stationary noise is still a challenging task.

In recent DNN-based methods, statistical assumptions about the distribution of noisy and clean \ac{STFTM} are typically dropped, while the minimum \ac{MSE} (MMSE) objective becomes a loss function for which a DNN optimizes by stochastic gradient descent. One of the most popular loss functions has been the \ac{MSE} between clean and enhanced \ac{STFTM}
\begin{equation}\label{mse}
L(\vec{G}; \vec{S}, \vec{X}) = \text{mean}(||\vec{S}-\vec{G}\odot\vec{X}||^2_2),
\end{equation}
where $\vec{A}$ denotes $|A[t,k]|$ in vector form, and $\odot$ is the element-wise product. A competitive method proposed recently \cite{valin2018hybrid} estimates the optimal gain function of a smoothed energy contour using a \ac{RNN} and interpolates the spectral details by pitch filtering. Experiments \cite{valin2018hybrid,Reddy2019} report strong objective and subjective speech quality from enhanced speech produced by this RNNoise system.

\section{Proposed Methods}
\label{sec:method}

\subsection{Feature representation}

Selecting appropriate features and normalization is important for successfully training a DNN. We consider two basic features in \ac{STFTM} and \ac{LPS}, and apply global, \ac{FD}, and \ac{FI} normalization, respectively, to train our network.

The \ac{STFTM} used in all our systems is computed based on a 32~ms Hamming window with 75\% overlap between frames and a 512-point discrete Fourier transform. The \ac{LPS} is taken with the natural logarithm and floored at -120 dB, i.\,e.,
\begin{equation}
    f_{LPS}(|X[t,k]|) = \log(\max(|X[t,k]|^2, 10^{-12}))
\end{equation}
We explore three types of normalization to be each individually combined with either \ac{STFTM} or \ac{LPS} mentioned above. First, we consider global normalization, in which case each frequency bin is standardized by its mean and standard deviation accumulated from a training set:
\begin{equation}
    g_{G}(f(|X[t,k]|)) = \left[f(|X[t,k]|)-\mu_{f(x)}[k]\right] / \sigma_{f(x)}[k]
\end{equation}
Second, we consider online \ac{FD} mean and variance normalization, in which case the running mean and variance are smoothed by a decaying exponential:
\begin{align}
&\mu_{f(x)}[t,k] = c \, \mu_{f(x)}[t-1,k] + (1-c) \, f(|X[t,k]|)\\
&\sigma^2_{f(x)}[t,k] = c \, \sigma^2_{f(x)}[t-1,k] + (1-c) \, f(|X[t,k]|)^2\\
& g_{FD}(f(|X[t,k]|)) = \frac{f(|X[t,k]|)-\mu_{f(x)}[t,k]}{\sqrt{\sigma^2_{f(x)}[t,k]-\mu_{f(x)}^2[t,k]}}
\end{align}
where $c=\text{exp}(-\Delta t/\tau)$, $\Delta t$ is the frame shift in seconds (8 milliseconds in our setting), and $\tau$ is a time constant that controls the adaptation speed. The idea is that the normalized spectra will facilitate the recurrent neural network learning long-term patterns. Finally, we also have the \ac{FI} online normalization, in which case the mean and variance from each frequency are averaged and applied to all frequencies. This method retains the relative dynamics across frequency bins, but might pose a more challenging learning task to the learning machine. In all our experiments apart from the feature experiments, we use \ac{FD} online normalization with $\tau=3s$.
\subsection{Learning machine}
Our learning machine that takes in one frame of noisy speech spectra and outputs one frame of magnitude gain function is based on the \ac{GRU} \cite{cho2014learning}. \ac{GRU}s are preferred over \ac{LSTM}\cite{hochreiter1997long} given their computational efficiency and superior performance in real-time SE tasks \cite{Reddy2019}. We stack three GRU layers followed by a fully-connected (FC) output layer with sigmoid activation to predict the gain function $G[t,k]$.

It is worth mentioning that we do not apply convolution layers as often done in other related work \cite{tan2018convolutional,zhao2018convolutional} because of the relatively arbitrary process involved in choosing the amount of frequency span and filter taps. Previous studies \cite{liu2014experiments} have shown that a na\"ive convolution layer applied on past and present input noisy frames did not improve objective quality of enhanced speech. Instead, we explore the temporal modeling capability of the network by training with sequences of different lengths, features, and loss functions.


%
\subsection{Loss functions}
We use three loss functions to train our system. First, we use the regular \ac{MSE} between clean and enhanced \ac{STFTM} in \eqref{mse}. To obtain a better control of the loss, we propose to separate the error into speech distortion and noise reduction terms
\begin{align}
&L_\text{speech} = \text{mean}(||\vec{S}_{\text{SA}}-(\vec{G} \odot \vec{S})_{\text{SA}}||^2_2)\\
&L_\text{noise} = \text{mean}(||\vec{G} \odot \vec{N}||^2_2)
\end{align}
where subscript SA denotes a subset of frames where speech is active. In our experiments, we adopted a simple energy-based frame-level voice activity detector operating on the power spectra of clean utterances. The short-time speech energy is accumulated between 300 Hz and 5000 Hz and smoothed over 3 frames by a moving-average filter. Finally, a frame is decided to be voiced above a threshold of 30 dB below the peak energy of the whole utterance.

As the estimated gain approaches all-pass, the speech distortion error is minimized and the noise error is maximized, and vice versa. Therefore, we can control the relative importance of speech distortion to noise reduction with a \textit{fixed-weighted} loss, 
\begin{equation}
\label{eq:weighted_loss}
L(\vec{G}; \vec{S}_{\text{SA}}, \vec{N}) = \alpha L_\text{speech} + (1-\alpha)L_\text{noise},
\end{equation}
where $\alpha$ is a constant in range [0, 1]. We notice a similar loss has been developed independently and termed two-component loss (2CL) \cite{xu2019components}. Next, we discuss an extension to this fixed weighting.

\begin{figure}[tb]
  \includegraphics[width=\columnwidth]{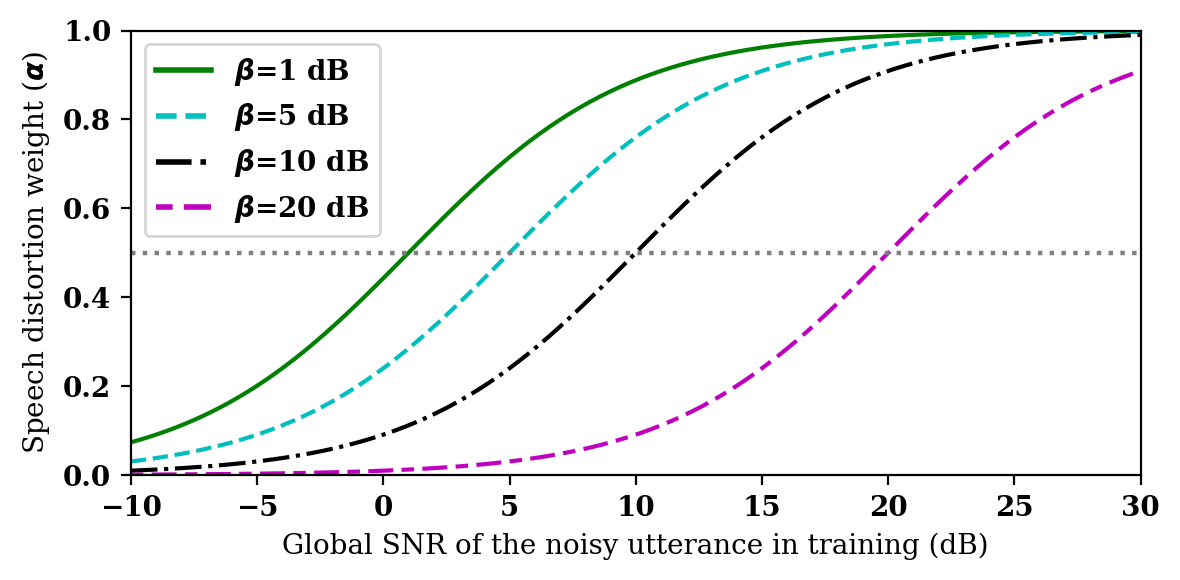}
  \caption{Selected \textit{SNR-weighted} speech distortion weighting. Horizontal line marks equal weighting of $L_{\text{speech}}$ and $L_{\text{noise}}$.}
  \label{fig:weight}
\end{figure}

In classical speech enhancement literature, the suppression rule is often adapted based on the \ac{SNR} \cite{esch2009,braun2015}. Specifically, suppression should be limited at high \ac{SNR} to avoid artifacts, and be aggressive at low \ac{SNR}. Motivated by this principle, our second \textit{SNR-weighted} loss adjusts $\alpha$ in \eqref{eq:weighted_loss} using the global \ac{SNR} of each utterance 
\begin{equation}
\alpha =  \frac{\text{SNR}}{\text{SNR} + \beta},
\end{equation}
where SNR $=||\vec{S}||^2_2/||\vec{N}||^2_2$ and $\beta$ is a constant. Note that \resizebox{0.3\columnwidth}{!}{$d\alpha/d[10\text{log}_{10}(\text{SNR})]$} is maximized when $\text{SNR} = \beta$. In this way, $\beta$ controls the global \ac{SNR} at which a fixed amount of deviation would cause the maximum drift in speech distortion weighting. Furthermore, $\beta$ also indicates the global \ac{SNR}, where the two loss terms are equally weighted. We illustrate this in Fig.~\ref{fig:weight}.

\begin{figure}[tb]
 \includegraphics[width=\columnwidth]{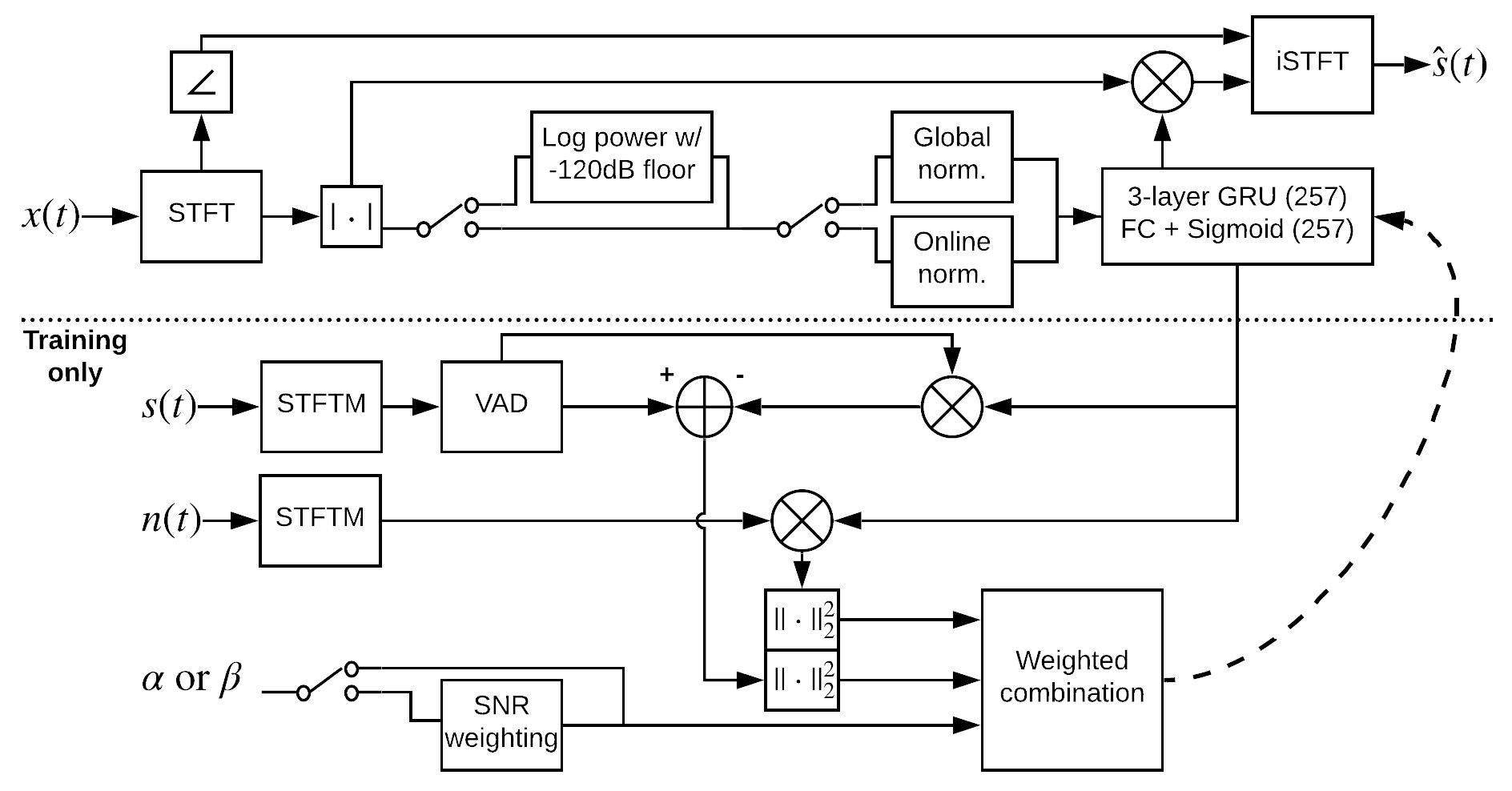}
 \setlength{\abovecaptionskip}{-5pt}
 \setlength{\belowcaptionskip}{-5pt}
 \caption{Flow diagram of the proposed system.}
 \label{fig:flow}
\end{figure}

The proposed method is depicted in a flow diagram shown in Fig.~\ref{fig:flow}. During training, both clean speech and noise are required for computing the weighted loss. The trained model enhances the noisy \ac{STFTM} one frame at a time and utilizes the noisy phase for reconstructing the enhanced speech waveform.
\section{Experimental Results and Discussions}
\label{sec:exp}
\subsection{Corpora~\& Experimental setup}
We train and evaluate all DNN-based systems using a dataset synthesized from publicly available speech and noise corpus using the MS-SNSD dataset \cite{Reddy2019} and toolkit \footnote{https://github.com/microsoft/MS-SNSD}. 14 diverse noise types are selected for training, while samples from 9 noise types not included in the training set are used for evaluation. Our test set includes challenging and highly non-stationary noise types such as munching, multi-talker babble, keyboard typing,~\emph{etc}. All audio clips are resampled to 16 kHz. The training set consists of 84 hours each of clean speech and noise while 18 hours (5500 clips) of noisy speech constitute the evaluation set. All speech clips are level-normalized on a per-utterance basis, while each noise clip is scaled to have one of the five global SNRs from \{40,30,20,10,0\} dB. During the training of all DNN-based systems described below, we randomly select an excerpt of clean speech and noise, respectively, before mixing them to create the noisy utterance. 

We performed a comparative study of proposed methods with three baselines based on several objective speech quality and intelligibility measures and subjective tests. Specifically, we include \ac{PESQ} \cite{rix2001perceptual}, \ac{STOI}\cite{taal2010short}, \ac{CD}, and \ac{SI-SDR}~\cite{le2019sdr} for objective evaluation of enhanced speech in time, spectral, and cepstral domains. We conducted a subjective listening test using a web-based subjective framework presented in \cite{Reddy2019}. Each clip is rated with a discrete rating between 1 (very poor speech quality) and 5 (excellent speech quality) by 20 crowd-sourced listeners. Training and qualification are ensured before presenting test clips to these listeners. The mean of all 20 ratings is the \ac{MOS} for that clip. We also removed obvious spammers who consistently selected the same rating throughout the MOS test. Our subjective test complements the other objective assessments, thus providing a balanced benchmark for evaluation of studied noise reduction algorithms.

We compare our proposed methods with three baseline methods. We used a classical enhancer, which is a slightly optimized implementation of the MMSE log-spectral amplitude (LSA) estimator \cite{ephraim1985speech} described in \cite{tashev2009sound}. DNN-based baselines include the improved RNNoise ($\text{RNNoise}_\text{I}$) \cite{Reddy2019} and a RNN ($\text{RNNoise}_{257}$) that replicates the network architecture of RNNoise \cite{valin2018hybrid} but operates on 257-point spectra, is trained on \eqref{mse}, and does not have the originally proposed post-processing component. $\text{RNNoise}_{257}$ realizes a system with a comparable number of parameters as the proposed methods. 

In the next section, we discuss the impact of feature normalization and training on various sequence lengths on the objective quality of enhanced speech. Then, we explore the optimal weighting for the proposed \textit{fixed-weighted} and \textit{SNR-weighted} loss functions. Finally, we compare the subjective and objective quality of enhanced speech produced by our systems to several competitive online methods.
\begin{table}[tb]
  \centering
  \caption{Effect of sequence lengths in a one-minute minibatch.}
  \label{tab:seqlen}
  \begin{tabular}{c|c|c|c|c}
    Length (s) &SI-SDR (dB)&CD&STOI (\%)&PESQ (MOS)\\
    \hline
    
    1&13.7&3.78&90.1&2.58 \\
    2&13.7&3.80&90.3&2.57 \\
    5&14.1&3.72&90.5&2.59 \\
    10&14.1&3.73&90.7&2.64 \\
    20&14.0&3.73&90.6&2.64 \\
  \end{tabular}
  \vspace{-4mm}
\end{table}
\begin{table}[tb]
  \centering
  \caption{Subjective MOS from 5500 clips and 20 ratings per clip.}
  \label{tab:sub}
  \begin{tabular}{l|l}
    Method & MOS (mean $\pm$ std.) \\
    \hline
    Noisy &2.63 $\pm$ 0.03 \\
    $\text{RNNoise}_\text{I}$ \cite{Reddy2019} &3.26 $\pm$ 0.03 \\
    Proposed ($\alpha=0.05$)&3.93 $\pm$ 0.03 \\
    Proposed ($\alpha=0.1$)&3.92 $\pm$ 0.03 \\
    Proposed ($\alpha=0.2$)&3.74 $\pm$ 0.03 \\
    Proposed ($\alpha=0.3$)&3.65 $\pm$ 0.03 \\
  \end{tabular}
\end{table}
\begin{table}[tb]
  \centering
  \caption{Comparison of objective metrics with baseline online SE systems. Refer to text for details about each setup.}
  \label{tab:obj}
  \begin{tabular}{c|c|c|c|c|c}
    Method&\# Param.&SI-SDR&CD&STOI&PESQ\\
          &         & (dB) & & (\%) & (MOS)\\
    \hline
    Noisy&--&9.81&4.56&88.0&2.22 \\
    LSA \cite{ephraim1985speech,tashev2009sound}&--&6.10&4.64&84.7&2.33 \\
    $\text{RNNoise}_\text{I}$ \cite{Reddy2019} &61.2 K&10.4&3.83&88.0&2.55 \\
    $\text{RNNoise}_{257}$ &2.64 M&13.0&3.88&89.3&2.56 \\
    $\text{Proposed}_{0.35}$&1.26 M&\bf{14.3}&\bf{3.83}&\bf{90.7}&\bf{2.65} \\
    Wiener&Oracle&20.5&2.13&98.1&3.82 \\
  \end{tabular}
  \vspace{-4mm}
\end{table}
\begin{figure*}[h]
  \includegraphics[width=\textwidth,height=0.7\columnwidth]{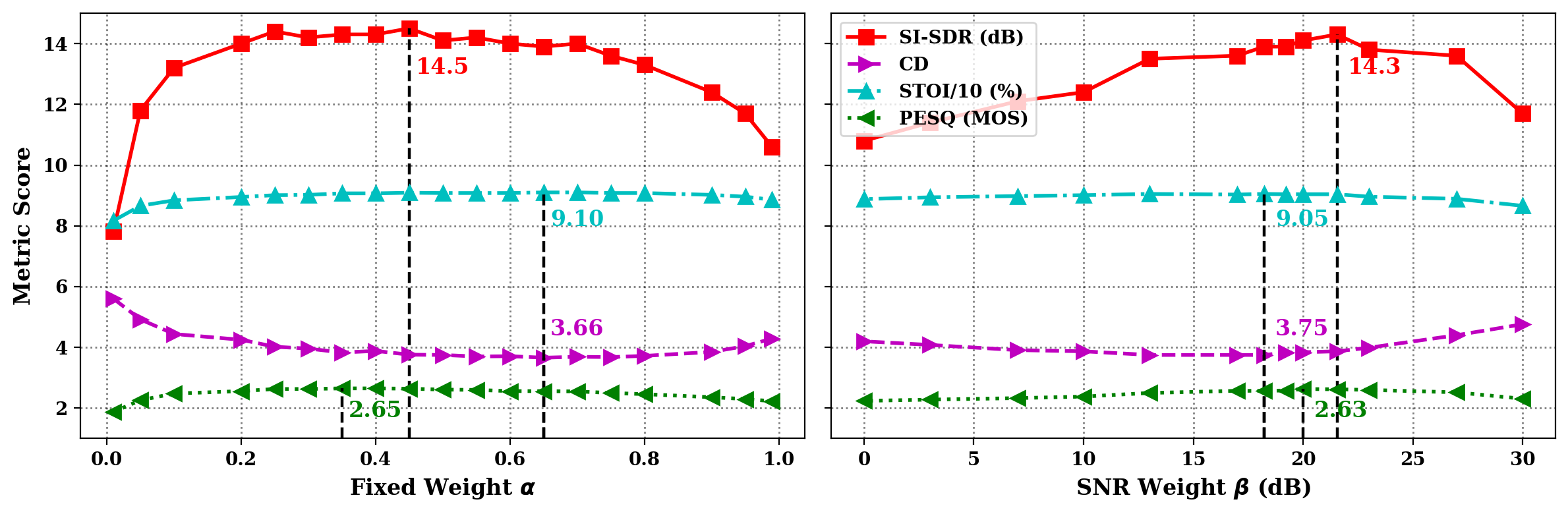}
  \caption{Effect of fixed weighting and SNR weighting on objective speech quality and intelligibility measures. Black dashed vertical lines indicate the optimal coefficient for each metric. Note that the optimal points coincide for STOI and CD at $\alpha = 0.65$ and $\beta=18.2$ dB.}
  \label{fig:weighting}
\end{figure*}
\vspace{-2mm}
\subsection{Results~\& Discussions}
\vspace{-1mm}
We want to evaluate how training with long or short sequences affects temporal modeling in the RNN. Although long sequences are expected to help deal with long-term noise patterns, it might also potentially degrade speech that is only short-term stationary. Table~\ref{tab:seqlen} summarizes this impact of sequence lengths on objective speech quality. For each setting, we adjust the number of sequences in a minibatch so that one batch always contains one minute of noisy speech. We observe a noticeable improvement in performance as each segment increases to 5 seconds, beyond which the improvement starts to diminish. We do not show the result for the feature test due to space limitation, but overall there is little difference between all normalized variants of \ac{STFTM} and \ac{LPS} features, while no normalization results in degradation. In general, we recommend \ac{FD} online normalization due to its invariance to varying signal levels. We also suggest using segments that are no less than 5-second long each during training.



The effect of speech distortion weighting is shown in Fig.~\ref{fig:weighting}, where $\alpha$ or $\beta$ are changed to search the optimal points for each objective measure. Curiously, only \ac{STOI} and \ac{CD} agree on the same coefficient in both cases, while both \ac{PESQ} and \ac{SI-SDR} suggest smaller weight on speech distortion. The optimal SNR weights for all metrics are concentrated around 20 dB, meaning that the speech distortion weight should only rapidly increase when the noisy signal is relatively clean. Overall, a fixed weighting is slightly better than SNR weighting in all metrics.

During experiments, we notice that even though our systems trained on \ac{MSE} (\emph{e.g.} row 4 in Table \ref{tab:seqlen}) could achieve similar objective measures compared to those trained on the proposed weighted losses \eqref{eq:weighted_loss}, the corresponding subjective quality of systems trained on the weighted loss is a lot better. The most noticeable improvement of systems trained on our loss functions, especially with small $\alpha$, is that the estimated gain function is much more frequency-selective than systems trained on regular \ac{MSE}, resulting in higher noise suppression, especially at high SNRs. To testify this, we present the result of the online subjective listening test in Table \ref{tab:sub}. Not only did all our selected systems significantly outperform the improved RNNoise ($\text{RNNoise}_\text{I}$) trained on \ac{MSE} presented in \cite{Reddy2019}, we were surprised that the listening test subjects preferred a rather low setting for the speech distortion weight $\alpha$. This trend is mispredicted by all objective measures as well as the authors' subjective preference of about $\alpha= 0.35$. We observed noticeable speech distortion as $\alpha$ goes below 0.35, while noise became more suppressed. It is evident that more detailed investigations are required in future work to shed more light on speech distortion and noise reduction preferences for different groups of listeners.

Finally, we report the objective evaluation from each baseline method, the noisy reference, and an oracle Wiener filter as upper bound in Table \ref{tab:obj}. The selected system from our method is trained using fixed speech distortion weighting with $\alpha=0.35$, which we believe strikes a good balance between speech distortion and noise reduction. Although this setup might not be the most preferred for human listeners, it can be easily tuned to different applications. It is nevertheless important to show that it outperforms all tested classical or DNN-based methods in all objective metrics.

\vspace{-2mm}
\section{Conclusions}
\vspace{-2mm}
\label{sec:con}
In this paper, we proposed and evaluated a real-time speech enhancement approach based on a compact recurrent neural network trained with a simple MSE-based speech distortion weighted loss function. We show the impact of various feature normalization techniques and sequence lengths on the objective quality of enhanced speech. We also demonstrate how to control the amount of speech distortion with fixed-weighted and SNR-weighted coefficients in the loss function. Both objective and subjective tests show that our method outperforms other competitive online methods. In the future, we will explore time-varying speech distortion weighting and its influence on subjective and objective speech quality.


%
%
%

\pagebreak

\bibliographystyle{IEEEbib-abbrv}
\bibliography{refs}

\end{document}